\documentclass[%
 reprint,
 amsmath,amssymb,
 aps,
]{revtex4-2}

\usepackage{graphicx}
\usepackage{dcolumn}
\usepackage{bm}

\begin{document}

\preprint{APS/123-QED}

\title{On-chip picosecond synchrotron pulse shaper}

\author{Jian Zhou,$^{1,2,\dagger}$ Jinxing Jiang,$^{1,\dagger}$ Donald A. Walko,$^{1}$ Dafei Jin,$^{2}$ Daniel López,$^{3}$ David A. Czaplewski,$^{2}$ and Jin Wang$^{1}$ }

\altaffiliation{wangj@anl.gov $^{\dagger}$These authors contribute equally.}

\affiliation{%
 $^{1}$Advanced Photon Source, Argonne National Laboratory, 9700 South Cass Avenue, Argonne, 60439, Illinois, USA
}%

\affiliation{%
 $^{2}$Center for Nanoscale Materials, Argonne National Laboratory, 9700 South Cass Avenue, Argonne, 60439, Illinois, USA
}%

\affiliation{%
 $^{3}$Materials Research Institute, Penn State University, University Park, PA 16802, USA
}%

\date{\today}

\begin{abstract}
Synchrotrons are powerful and productive in revealing the spatiotemporal complexities in matter. However, X-ray pulses produced by the synchrotrons are predetermined in specific patterns and widths, limiting their operational flexibility and temporal resolution. Here, we introduce the on-chip picosecond synchrotron pulse shaper that shapes the sub-nm-wavelength hard X-ray pulses at individual beamlines, flexibly and efficiently beyond the synchrotron pulse limit. The pulse shaper is developed using the widely available silicon-on-insulator technology, oscillates in torsional motion at the same frequency or at harmonics of the storage ring, and manipulates X-ray pulses through the narrow Bragg peak of the crystalline silicon. Stable pulse manipulation is achieved by synchronizing the shaper timing to the X-ray timing using electrostatic closed-loop control. Tunable shaping windows down to 40 $ps$ are demonstrated, allowing X-ray pulse picking, streaking, and slicing in the majority of worldwide synchrotrons. The compact, on-chip shaper offers a simple but versatile approach to boost synchrotron operating flexibility and to investigate structural dynamics from condensed matter to biological systems beyond the current synchrotron-source limit.  
\begin{description}
\item[Keywords]
Synchrotron, MEMS, optics, ultrafast X-rays

\end{description}
\end{abstract}

\maketitle

For decades, the use of synchrotrons has been revolutionizing science and technology with an impact on many aspects of our lives. Dozens of synchrotron X-ray sources around the world are serving more than 55,000 users annually across a diverse scientific community \cite{cho2020x}, enabling the study of matter in short scales of length and time, such as probing protein structures and folding dynamics \cite{rasmussen2011crystal,hekstra2016electric}, physical and chemical information and processes \cite{griffith2018niobium,cunningham2019keyhole,sun2012three,macphee2002x}, and mantle constitution and activities \cite{greaux2019sound,chanyshev2022depressed}. At a synchrotron source, electrons circulate inside the storage ring at nearly the speed of light and generate intense and periodic X-ray pulses with small wavelength and short duration, which are essential for the study of matter with high spatiotemporal resolution. Compared with free-electron lasers which could produce brighter and shorter X-ray pulses, synchrotron sources have considerably larger beamline capacity, wide photon energy tunability, outstanding beam stability, high average flux, and broad field of view \cite{willmott2019introduction,deng2021experimental}. 

Despite the impressive developments in synchrotron techniques, there remain limitations in producing favorable X-ray pulses to fulfill various needs at different beamlines \cite{willmott2019introduction}. To provide specific pulses, there are mainly two types of approaches: internal configuration of the electron bunches \cite{schoenlein2000generation,beaud2007spatiotemporal,zholents1996femtosecond,holldack2014single,martin2011experience}, and the use of external X-ray devices to shape the pulses \cite{forster2015phase,osawa2017development,grigoriev2006subnanosecond,larsson1998ultrafast,gaal2014ultrafast,chen2019ultrafast}. Although synchrotron X-ray pulses can be modified by manipulating the electron bunches such as slicing the electron bunch using laser pulses or operating the storage ring in specific timing modes, these approaches are usually complex and costly, or come at the expense of pulse characteristics and beamline operating compatibility \cite{schoenlein2000generation,beaud2007spatiotemporal,zholents1996femtosecond,holldack2014single,martin2011experience}. Alternatively, X-ray pulses can be adapted at the individual beamlines with external X-ray devices such as optical choppers and Bragg switches \cite{forster2015phase,osawa2017development,grigoriev2006subnanosecond,larsson1998ultrafast,gaal2014ultrafast,chen2019ultrafast}. Despite great efforts over the past decades, none of the X-ray modification devices possess the capability to efficiently shape the individual synchrotron pulses. 

Here, we demonstrate an on-chip picosecond synchrotron pulse shaper that enables the delivery of sub-nm-wavelength hard X-ray pulses flexibly and efficiently at various individual beamlines with the potential to go beyond the intrinsic pulse characteristics of those synchrotrons. A tunable shaping window down to 40 $ps$ is demonstrated, allowing flexible X-ray manipulations including pulse picking, streaking, and slicing in the majority of synchrotrons across the globe. The simple on-chip shaper offers possibilities to enhance synchrotron operating flexibility and to investigate structural dynamics beyond the pulse limits of those synchrotrons.

\begin{figure*}[htp]
	\centering 
	\includegraphics[page=1, trim = 0.7in 2.05in 5.3in 1.7in, clip, width=0.94\textwidth]{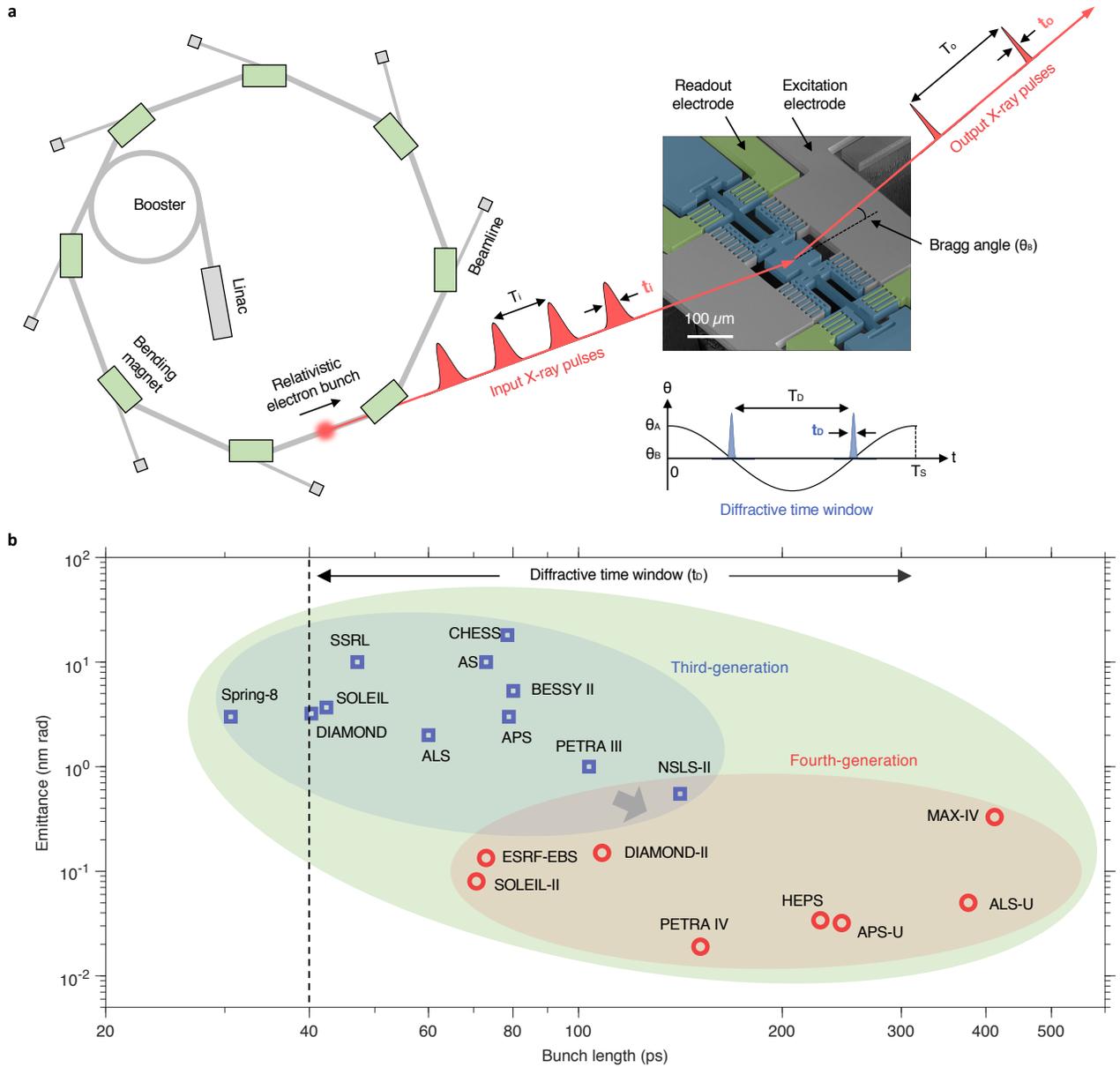}
	\caption{\textbf{$\vert$ Shaping the synchrotron radiations flexibly at individual beamline.} \textbf{a}, Schematic of X-ray pulse shaping with the ultrafast shaper on-chip. The MEMS shaper is composited of single crystal silicon ($t=25\;\mu m$), diffracts X-ray pulses through the narrow Bragg peak of material. When operated in ultrafast periodic torsional oscillation, the shaper can not only modulate the X-ray bunch pattern ($T_i\to  T_o$), but also shape the X-ray bunch length ($t_i\to  t_o$). \textbf{b}, Bunch length (FWHM) of worldwide synchrotron light sources in standard operation mode. The shaper allows tunable bunch picking, streaking, and shaping for the majority of light sources.
	}
	\label{fig.01}
\end{figure*}

The concept of synchrotron pulse shaping with the dynamic on-chip shaper is shown schematically in Fig.\ref{fig.01}. The ultrafast shaper modulates the X-ray bunch pattern ($T_i\to  T_o$) and shapes the X-ray bunch length ($t_i\to t_o$) in an individual beamline. The shaper is a photonic microelectromechanical system (MEMS) consisting of three functional areas: the excitation electrodes, the readout electrodes, and the dynamic torsional body (Fig.\ref{fig.01}a). It is fabricated from single crystal silicon with thickness 25 $\mu m$, which diffracts X-ray pulses through a narrow peak ($\Delta\theta_B$ in millidegrees) at the Bragg angle, $\theta_B$. The rectangular diffractive mirror in the torsional body has a surface area of 100 $\mu m$ $\times \;$70 $\mu m$, and is supported by a pair of torsional beams. When the torsional body is electrostatically excited into oscillation using the comb drive fingers, the narrow Bragg peak forms a diffractive time window for dynamic X-ray manipulation \cite{chen2019ultrafast}. Unlike macroscopic optics which usually have no or limited speed of motion (e.g. monochromator and optical chopper), the microscopic optics can be operated with ultrafast angular velocity, providing the opportunity to achieve a diffractive time window ($t_D$) even narrower than the bunch length of the synchrotron that may vary from tens to hundreds of picoseconds (Fig.\ref{fig.01}b). Thereby, the shaper can enable spatiotemporal experiments to be performed with a resolution beyond the synchrotron pulse limit, without the need of complex and costly modifications of those synchrotrons and their corresponding detrimental effects to the pulse characteristics or reduced beamline operating compatibility.

\begin{figure*}[htp]
	\centering 
	\includegraphics[page=1, trim = 0.8in 2.4in 5.3in 1.7in, clip, width=0.94\textwidth]{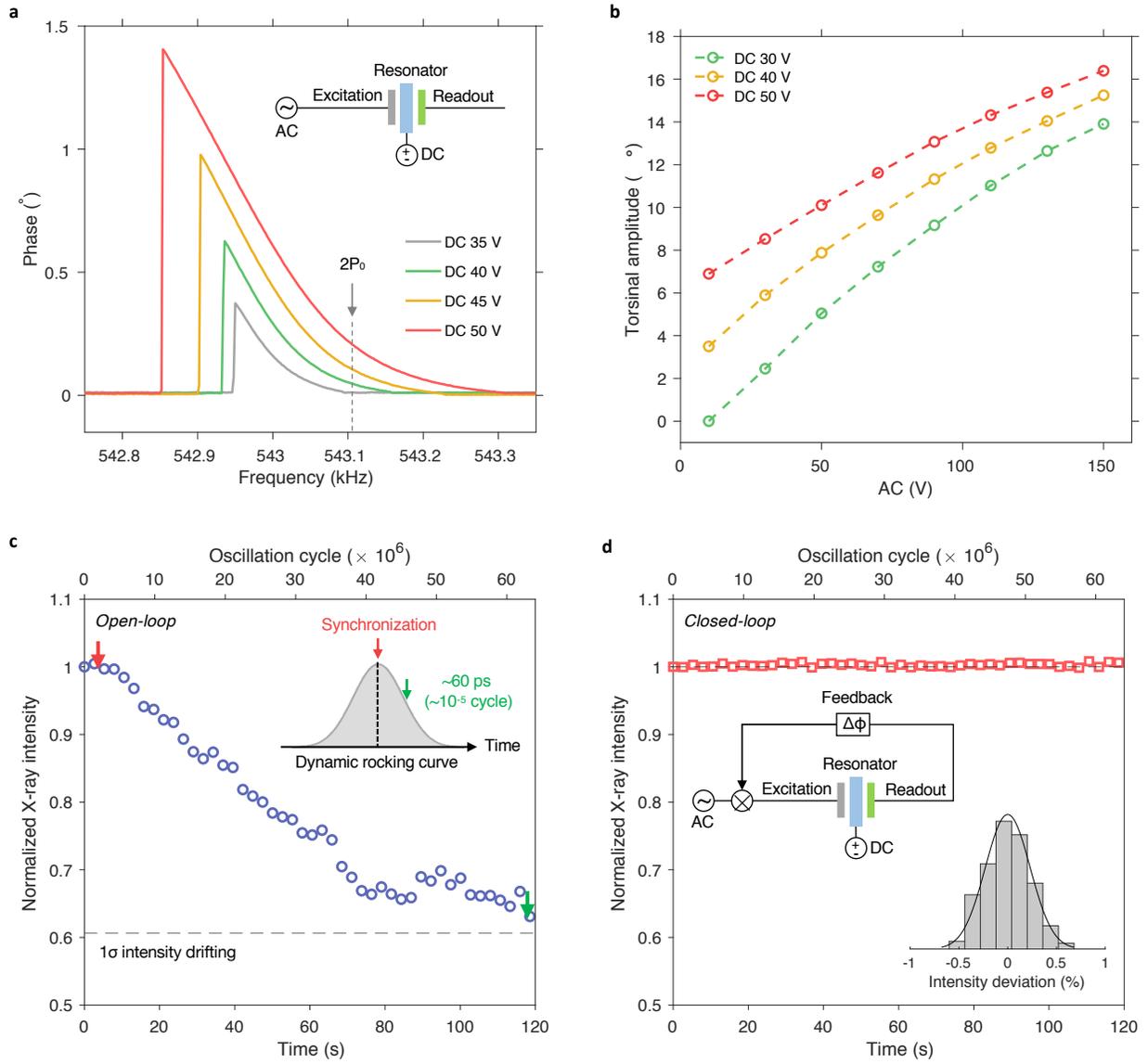}
	\caption{\textbf{$\vert$ Synchronized hard X-ray pulse manipulation.} \textbf{a}, Tuning curves of the MEMS shaper. The insert figure shows the circuit diagram. Phase readouts were measured at 10 $V$ AC excitation voltage, under vacuum condition (1 $Pa$). Device can be excited at 2$P_0$ (543106 $Hz$), where $P_0$ is the operation frequency of the APS. \textbf{b}, Tunable torsional amplitude at 2$P_0$ excitation frequency. \textbf{c}, Open-loop synchronization stability. The input X-rays are generated in the APS 24-bunch mode (8 $keV$ energy, 78.7 $ps$ bunch length). The output X-rays were measured with a fast avalanche photodiode at 8.8$^\circ$ torsional amplitude, where the dynamic rocking curve has a width of 127 $\pm$ 2 $ps$ (see Fig.\ref{fig.S01}c). \textbf{d}, Closed-loop synchronization stability. The insert figures show the closed-loop circuit diagram and the intensity deviation.       }
	\label{fig.02}
\end{figure*}

To efficiently manipulate the periodic x-ray pulses produced by a light source, the on-chip shaper was designed with the following characteristics: 1) a tailorable frequency to precisely match the periodic x-ray pulses; 2) an unprecedented combination of efficiency and temporal resolution for x-ray pulse shaping; 3) a broadly tunable window to allow user-defined flexible X-ray manipulations such as pulse picking, streaking, and slicing; and 4) long-term dynamic operation stability. For this work, all measurements were performed at the Advanced Photon Source (APS) in the 24-bunch standard operating mode. The input X-ray pulse length is 78.7 $ps$ in full width half maximum (FWHM), with a photon energy of 8 $keV$. The frequency of a relativistic electron bunch looping around the APS storage ring of 1.1 $km$ circumference is $P_0=271553\;Hz$. Further experimental details are provided in the Methods section.

The dynamic on-chip shaper has a precisely tailored resonance frequency, to synchronize with the characteristic frequency of the APS storage ring. Because of the torsional operation of the shaper, the body passes by the equilibrium position twice for each cycle of oscillation, making the operating frequency 2$P_0$ (Fig.\ref{fig.02}a). To drive the resonator, an AC voltage is applied to the excitation electrodes, and the torsional body is biased with a DC voltage as shown in the insert figure of Fig.\ref{fig.02}a. The phase response of the resonator is measured from the readout electrodes \cite{antonio2012frequency}. 

Note that the dynamic shaper has an extremely narrow bandwidth. Directly fabricating a device with a resonance at 271553 $Hz$ is beyond the current microfabrication capability and thus requires tuning of the resonant frequency. To make the shaper frequency match with the storage ring, we designed the resonator with a torsional resonance frequency slightly higher (5$\%$) than $P_0$ by using finite element analysis-assisted design. The as-fabricated device demonstrated a torsional resonance frequency 3$\%$ higher than $P_0$ due to the inaccuracy of the microfabrication process. By thinning the torsional beams in the thickness direction using $Ga^+$ focus ion beam milling, the resonant frequency was brought close to $P_0$ \cite{,zhou2020approaching}. The torsional beams, after thinning, can be seen in the scanning electron microscope image in Fig.\ref{fig.01}a, where the mechanical beams are a slightly different color from the other parts of the torsional body. Detailed shaper design, fabrication, and characterization processes are provided in the Methods sections. Frequency tuning steps can be performed to match the resonance frequency of a shaper with other synchrotron sources. 

The shaper diffractive time window can be estimated as 
\begin{equation} \label{eq1}
t_D=\Delta\theta_B/\Omega
\end{equation}
where $\Delta\theta_B$ is the crystal rocking curve width and $\Omega$ is the angular speed of the torsional resonator when the X-ray pulse impinges on the diffractive MEMS element. A diffractive peak was observed around the Si (400) Bragg angle, $\theta_B$ of 34.803$^\circ$, with a static rocking curve width, $\Delta\theta_B$ of 0.0011 $^\circ$ in FWHM (see Fig.\ref{fig.S01}a). For a resonator with resonance frequency, $f$, and torsional amplitude, $\theta_A$, the maximum angular speed occurs at its equilibrium position, $\Omega_m=2\pi f\theta_A$. A larger torsional amplitude or a higher frequency results in a larger angular velocity and hence a smaller diffractive time window. By making the incident X-ray beam at the Bragg angle with the equilibrium position, the narrowest diffractive time can be estimated as $t_D=\Delta \theta_B / 2\pi f \theta_A$, which occurs periodically when the torsional body passes by the equilibrium position, as shown in Fig.\ref{fig.01}a. Note that the torsional body passes by the equilibrium position twice for each cycle of oscillation. Accordingly, the frequency of the output x-rays is 2$f$ ($f=P_0$), which has been confirmed experimentally as shown in the Fig.\ref{fig.S01}b.

When operating the microdevice at atmospheric pressures, air damping is the major source of energy dissipation. To efficiently drive the resonator into high speed motion, we operated the device in vacuum. Specifically, we reduced the gas pressure to make the resonator work in the ballistic regime, where the effect of air damping on the energy dissipation is ignorable \cite{schmid2016fundamentals}. This can be achieved when the mean free path length of the gas, $\lambda=k_B T/\sqrt{2} \pi d^2 p$ is larger than the representative physical length scale of the micromechanical structure, $L$, where $k_B$ is the Boltzmann constant, $T$ is absolute temperature, $d$ is the diameter of the gas particles, $p$ is the gas pressure, $L\sim 100\;\mu m$ is the length of the torsional mirror. For air at atmospheric pressure, the mean free path is approximately 70 $nm$. We estimated that the device will work in the ballistic regime when the pressure is less than 70 $Pa$. In practice, we operate the device in a pressure level of 1 $Pa$.  

\begin{figure*}[htp]
	\centering 
	\includegraphics[page=1, trim = 0.7in 2.5in 5.3in 1.7in, clip,  width=0.94\textwidth]{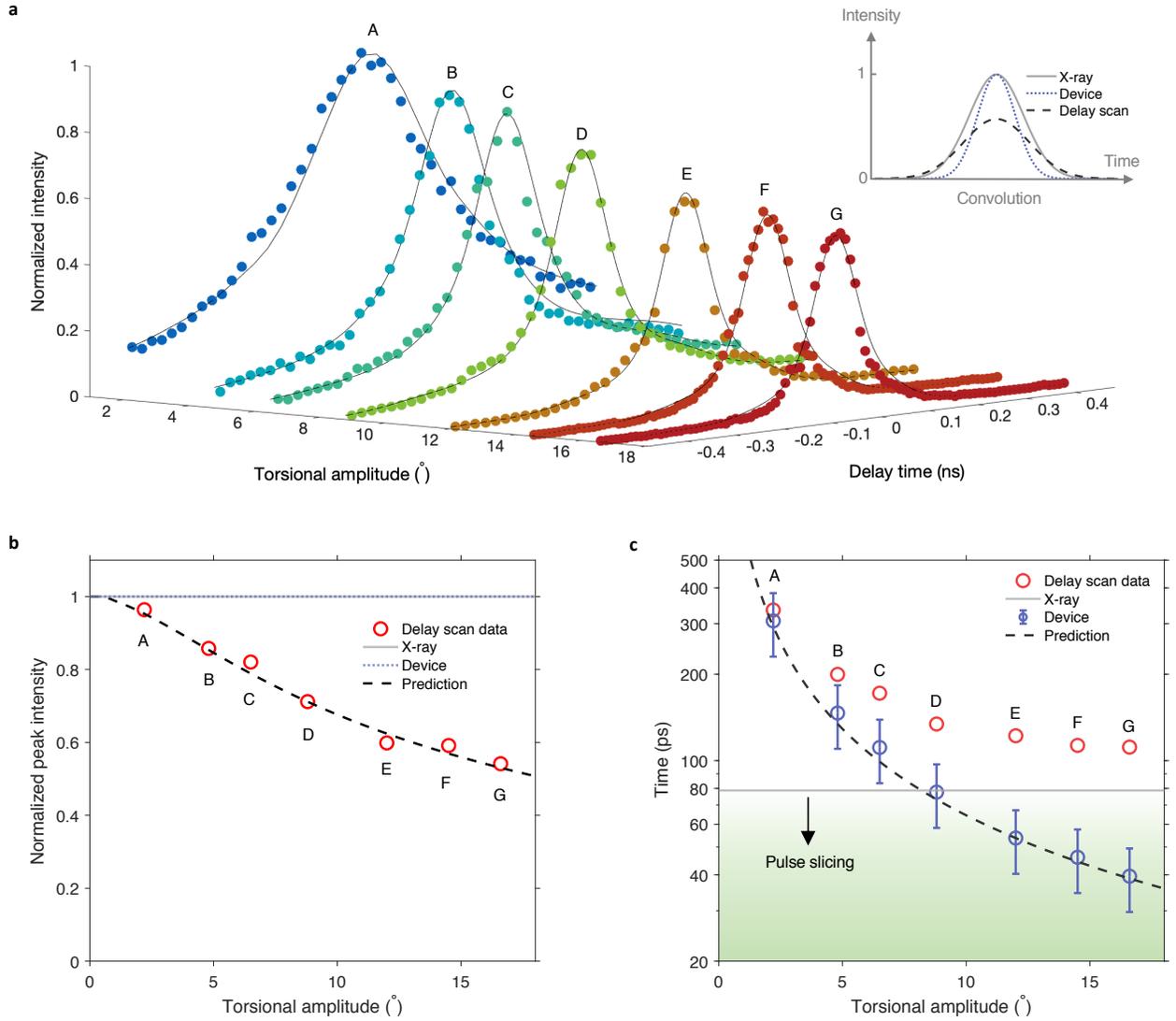}
	\caption{\textbf{$\vert$ X-ray pulse shaping.} \textbf{a}, Shaper dynamic rocking curves measured from the delay time scan in the APS 24-bunch mode. The dynamic rocking curve is the convolution of the X-ray pulse (width 78.7 $ps$) and the shaper diffractive time window ($t_D$), as illustrated in the insert figure. \textbf{b}, Normalized peak intensity from the delay time scan. \textbf{c}, Diffractive time window of the device measured from the delay time scan. Dash lines in b and c show the predictions according to equation \ref{eq1}. 
}\label{fig.03}
\end{figure*}

At 2$P_0$ excitation frequency, the torsional amplitude of the resonator can be tuned in a broad range of amplitudes from 2.2$^\circ$ to 16.7$^\circ$ by simply adjusting the voltage level (Fig.\ref{fig.02}b). The higher the AC or DC excitation voltage, the higher the torsional amplitude. The maximum angular speed approaches 2.8 $\times \;10^7\;^{\circ}/s$ when the resonator is driven to 16.7$^\circ$. According to equation \ref{eq1}, a wide range of diffractive time windows ranging from 293 $ps$ to 39 $ps$ are expected. This temporal tunability supports pulse picking, streaking, and slicing in the majority of worldwide synchrotrons as shown in Fig.\ref{fig.01}b. 

As a dynamic optics, operational stability is one of the most crucial parameters for periodic X-ray manipulation. For a resonator oscillating at $P_0$, the period is 3.68 $\mu s$. One part per million (ppm) of period drift is 3.68 $ps$ (corresponds to a phase shift of 0.00036$^\circ$). Therefore, stable manipulation of the dynamic X-ray pulses with the picosecond temporal resolution requires synchronizing the shaper timing to the X-ray timing at the scale of ppm cycling time. In practical applications, however, the operational stability of dynamic optics can be easily changed by different factors such as the temperature-induced frequency drifting in the resonator frequency and the low-frequency flicker noise in the electronics \cite{schmid2016fundamentals}. At 2$P_0$ excitation frequency, the phase noise of the electronic system is provided in Fig.\ref{fig.S02}. An example trace of the open-loop stability is shown in Fig.\ref{fig.02}c, where the dynamic optics loses synchronization with the periodic X-ray pulses at one second as indicated by the arrow. 

We designed the MEMS shaper with electrostatic feedback control to achieve long-term synchronized operation of the dynamic optics. The closed-loop circuit diagram is shown in the insert figure of Fig.\ref{fig.02}d. By tracking the resonator phase from the readout electrodes, we correct the phase drift of the resonator in relation to the x-ray timing in real time. Consequently, stable dynamic X-ray operation was achieved (Fig.\ref{fig.02}d), with a standard intensity deviation as low as 0.2$\%$. The close-looped control permits the practical application of the dynamic optics on-a-chip for stable X-ray manipulation robust to environmental disturbance and electronic drift.  

To resolve the shaper diffractive time window experimentally, we measured the dynamic rocking curve by varying the time delay of the torsional resonator in relation to the X-ray timing (Fig.\ref{fig.03}). Fig.\ref{fig.03}a shows a series of the dynamic rocking curves measured at different torsional amplitudes. A dynamic rocking curve is the convolution of the X-ray pulse (78.7 $ps$ width) and the shaper diffractive time window ($t_D$) as illustrated in the insert figure of Fig.\ref{fig.03}a. For an input X-ray pulse of fixed width, the narrower the diffractive time window, the less the peak intensity and width of the output X-ray pulse. In Fig.\ref{fig.03}b and \ref{fig.03}c, the dashed lines show the predictions. A predicted dynamic rocking curve at a torsional amplitude is the convolution of the X-ray pulse and the predicted diffractive time window according to equation \ref{eq1}. As shown in Fig.\ref{fig.03}b, we observe closely matched data between the experimental peak intensities of the delay scan traces with the predictions. From the time delay measurements, the resolved diffractive time window ($t_D$) ranges from 307 $\pm$ 77 $ps$ to 40 $\pm$ 10 $ps$ at torsional angle 2.2$^\circ$ to 16.7$^\circ$ (see Fig.\ref{fig.03}c, Fig.\ref{fig.S03}). The experimental diffractive time windows match closely with the prediction of equation \ref{eq1} from 293 $ps$ to 39 $ps$, suggests that the jitter noise of the system is low.

In this work, we realized the on-chip picosecond synchrotron pulse shaper that enables the delivery of hard X-ray pulses efficiently at individual beamlines beyond the intrinsic pulse limit of synchrotrons. With ultrafast X-ray optics, users can access narrower bunch lengths and higher brightness simultaneously in the next generation of synchrotron light sources, without the need of costly and complex modification of storage rings. For example, the shaper can slice the pulse width of APS-U and MAX-IV by a factor of 610$\%$ and 1030$\%$, respectively. The dynamic on-chip shaper offers the possibility to boost the synchrotron operating flexibility and to investigate structural dynamics in the condensed matter beyond the pulse limit of synchrotrons.  

\section*{Methods}
\textbf{Device design and fabrication}. The MEMS shaper consists of three components, the crystal diffractive element supported by the torsional beams, excitation electrodes, and readout electrodes. The material is a (001)-oriented silicon crystal of thickness 25 $\mu m$. The diffractive pad of the torsional body has a smooth area of 100 $\mu m$ $\times \;$70 $\mu m$, supported by a pair of torsional beams ($L=188\;\mu m$, $W=11.3\;\mu m$, $t=25\;\mu m$). Electrostatic comb-fingers are designed for torsional motion actuation and sensing. Finite element analysis was utilized to estimate the vibration mode and calculate the resonance frequency. Fabrication was performed at a commercial foundry MEMSCAP using the SOIMUMPS process.   

\textbf{Frequency tuning}. The resonance frequency of the device is precisely tuned by milling the torsional beams in the thickness direction with a $Ga^+$ ion-focused beam (FEI Nova Nanolab 600). The fractional frequency shift of the torsional resonator $\Delta f/f$ is approximately proportional to the relative change of the beam thickness $\Delta t/t$. To see this, the resonance frequency of a torsional resonator can be modeled as $f=(k/I)^{1/2}$, where $k$ is the stiffness of the torsional beams, $I$ is the moment of the inertia of the torsional body. Taking the derivative of this formula, the fractional frequency shift can be estimated as $\Delta f/f=0.5\Delta k/k$. For a pair of rectangular torsional beams, the torsional stiffness is $k=2GJ/L_e$, where $G$ is the modulus of rigidity, $L_e$ is the effective beam length, $J \approx \beta tw^3$ is the torsional constant, $\beta$ is a constant depending on $t/w$ \cite{,ugural2003advanced}. Therefore, the fractional frequency shift can be estimated as $\Delta f/f \approx \Delta t/t$. For a torsional resonator of $t\sim 25\; \mu m$ and $f\sim 272\;kHz$, the frequency tuning rate can be estimated as $\Delta f/\Delta t \approx 10.9\;Hz/nm$.  

\textbf{Electronic setup and closed-loop feedback}. The schematic of the electronic setup is shown in Fig.\ref{fig.S04}a. To manipulate the periodic X-rays, we use the APS RF signal (351.93 $MHz$) as the external reference for the lock-in amplifier (Zurich Instruments UHFLI), after dividing it by a factor of 4 with a frequency divider. Taking the APS signal as the reference, the lock-in amplifier generated a sinusoidal signal at 2$P_0$. After amplifying it with a voltage amplifier (Amplifier Research 75A250), this AC signal was applied to the excitation electrodes of the MEMS resonator. A bias voltage was applied to the torsional body using a DC power supply (Keysight E3634A). 

Due to the nonlinearity, the MEMS shaper cannot be driven to torsional mode directly at 2$P_0$. To drive the resonator at 2$P_0$ with the APS RF signal (Fig.\ref{fig.S04}a), we first excited the device to oscillate at 2$P_0$ using a function generator, by sweeping the frequency in the backward direction from 543.5 $kHz$ to 2$P_0$. Then we switched the excitation signal of the same frequency and amplitude from the function generator to the lock-in amplifier using a fast RF switch. 

The output signals from the readout electrodes were measured with the lock-in amplifier. The phase information was demodulated using the APS RF signals as the reference. For closed-loop feedback control, the real-time phase shift of the MEMS resonator in relation to the APS RF signal was compensated to the 2$P_0$-driven signal generated by the lock-in amplifier. 

For the device tuning curve measurement (Fig.\ref{fig.02}a) where synchronization was not required, we swept frequency backward in the desired frequency range using the lock-in amplifier. Instead of using the APS RF signal as the external reference, we used the internal oscillator of the lock-in amplifier as the reference to generate the AC excitation signal and demodulate the phase information of the MEMS resonator.

\textbf{Torsional amplitude characterization}. The torsional amplitude $\theta_A$ was characterized optically, as shown in Fig.\ref{fig.S04}b. By directing a laser beam to the torsional body, the reflected beam was projected to the monitoring screen aligned at distance $d$. When the torsional body is in periodic motion, the reflected beam forms a footprint of length 2$L_f$ in the monitoring screen. The torsional amplitude is calculated as $\theta_A=0.5tan^{-1}(L_f/d)$. 

\textbf{X-ray setup}. The schematic of the X-ray manipulation setup is shown in Fig.\ref{fig.S04}c. All measurements were performed at the standard 24-bunch mode of APS in beamline 7-ID. The input X-ray has a pulse length of 78.7 $ps$ in FWHM and a pulse interval of 153.4 $ns$. The incident X-ray pulses from the synchrotron were monochromatized to an energy of 8 $keV$ using a double-crystal monochromator. The X-ray beam was focused horizontally by a Kirkpatrick–Baez mirror, and adjusted to 10 $\times$ 10 $\mu m^2$ using slits. The device was housed in a vacuum chamber mounted on a high-precision diffractometer with an angular resolution of 3 $\times$ $10^{-5}$ degrees. 

We pumped the vacuum chamber using a vacuum pump (Edwards nXDS15i) and achieved a pressure level of 1 $Pa$. The incident X-ray beam was aligned to the center of the torsional mirror at the Bragg angle ($\theta_B=34.803^{\circ}$) in relation to the equilibrium surface of the mirror. We monitored the input x-ray intensity with an ionizing chamber. After passing the crystal analyzer (400), the diffracted X-rays were measured with an avalanche photodiode with a response time of 5 $ns$. While is not shown in the schematic figure, the photodiode signal was either sent to a scaler to acquire the X-ray intensity or to a high-speed digitizing oscilloscope (Yokogawa DLM4058) to record the real-time X-ray response. 

\textbf{X-ray measurement}. If the incident X-ray beam makes an angle $\theta_0$ with the equilibrium position of the mirror surface, the time dependence of the incident angle $\theta (t)$ during the oscillation can be described as $\theta (t)=\theta_0+\theta_A cos(2\pi ft+\Phi)$, where $\theta_A$ is the torsional amplitude, $f$ is the frequency, $\Phi$ is the phase of the shaper. The angular velocity of the shaper is given by $\Omega (t)=-\Omega_m sin(2\pi ft+\Phi)$, where $\Omega_m=2\pi f\theta_A$ is the maximum angular velocity that occurs at the equilibrium position at $\theta_0$. We aligned the incident beam at the Bragg angle, $\theta_0=\theta_B$ in relation to the equilibrium surface of the torsional mirror, such that the narrowest diffractive time window occurs periodically when the torsional body passes by the equilibrium position.

The static rocking curve as shown in Fig.\ref{fig.S01}a was measured when the device was not excited into torsional motion. We varied the surface of the torsional mirror in relation to the x-ray incident direction using the high-precision diffractometer. The diffractive X-rays were detected by the avalanche photodiode operated in photon-counting mode. Around the Si (400) Bragg angle $\theta_B=34.803^{\circ}$, we observed a narrow peak $\Delta \theta_B=0.0011^{\circ}$ in FWHM. 

The transient x-ray characterization as shown in the Fig.\ref{fig.S01}b was measured with the high-speed digitizing oscilloscope, when the device was static (OFF) or in torsional motion (ON). The device was synchronized to the periodic x-ray pulses under closed-loop feedback. The data were collected in the oscilloscope averaging mode (N=1024) with a sampling rate of 125 $GHz$. 

To resolve the shaper diffractive time window experimentally, we measured the dynamic rocking curve by varying the time delay of the torsional resonator in relation to the X-ray timing. The time delay was generated by adjusting the phase of the excitation signal using the lock-in amplifier. The measured dynamic rocking curve is the convolution of the X-ray pulse (78.7 $ps$ width) and the shaper diffractive time window.

\begin{acknowledgments}
This work was supported by the DOE/OS/BES/SUF Accelerator and Detector Research Program. This research used resources of the Advanced Photon Source and Center for Nanoscale Materials, U.S. Department of Energy (DOE) Office of Science user facilities at Argonne National Laboratory and is based on research supported by the U.S. DOE Office of Science-Basic Energy Sciences, under Contract No. DE-AC02-06CH11357. We thank R.T. Keane for lending us the broadband amplifiers.

\end{acknowledgments}

\bibliography{reference}

\providecommand{\noopsort}[1]{}\providecommand{\singleletter}[1]{#1}%
\begin{thebibliography}{26}%
\makeatletter
\providecommand \@ifxundefined [1]{%
 \@ifx{#1\undefined}
}%
\providecommand \@ifnum [1]{%
 \ifnum #1\expandafter \@firstoftwo
 \else \expandafter \@secondoftwo
 \fi
}%
\providecommand \@ifx [1]{%
 \ifx #1\expandafter \@firstoftwo
 \else \expandafter \@secondoftwo
 \fi
}%
\providecommand \natexlab [1]{#1}%
\providecommand \enquote  [1]{``#1''}%
\providecommand \bibnamefont  [1]{#1}%
\providecommand \bibfnamefont [1]{#1}%
\providecommand \citenamefont [1]{#1}%
\providecommand \href@noop [0]{\@secondoftwo}%
\providecommand \href [0]{\begingroup \@sanitize@url \@href}%
\providecommand \@href[1]{\@@startlink{#1}\@@href}%
\providecommand \@@href[1]{\endgroup#1\@@endlink}%
\providecommand \@sanitize@url [0]{\catcode `\\12\catcode `\$12\catcode
  `\&12\catcode `\#12\catcode `\^12\catcode `\_12\catcode `\%12\relax}%
\providecommand \@@startlink[1]{}%
\providecommand \@@endlink[0]{}%
\providecommand \url  [0]{\begingroup\@sanitize@url \@url }%
\providecommand \@url [1]{\endgroup\@href {#1}{\urlprefix }}%
\providecommand \urlprefix  [0]{URL }%
\providecommand \Eprint [0]{\href }%
\providecommand \doibase [0]{https://doi.org/}%
\providecommand \selectlanguage [0]{\@gobble}%
\providecommand \bibinfo  [0]{\@secondoftwo}%
\providecommand \bibfield  [0]{\@secondoftwo}%
\providecommand \translation [1]{[#1]}%
\providecommand \BibitemOpen [0]{}%
\providecommand \bibitemStop [0]{}%
\providecommand \bibitemNoStop [0]{.\EOS\space}%
\providecommand \EOS [0]{\spacefactor3000\relax}%
\providecommand \BibitemShut  [1]{\csname bibitem#1\endcsname}%
\let\auto@bib@innerbib\@empty
\bibitem [{\citenamefont {Cho}(2020)}]{cho2020x}%
  \BibitemOpen
  \bibfield  {author} {\bibinfo {author} {\bibfnamefont {A.}~\bibnamefont
  {Cho}},\ }\bibfield  {title} {\bibinfo {title} {X-ray source gets a 100-fold
  boost in brightness},\ }\href@noop {} {\bibfield  {journal} {\bibinfo
  {journal} {Science}\ }\textbf {\bibinfo {volume} {369}},\ \bibinfo {pages}
  {234} (\bibinfo {year} {2020})}\BibitemShut {NoStop}%
\bibitem [{\citenamefont {Rasmussen}\ \emph {et~al.}(2011)\citenamefont
  {Rasmussen}, \citenamefont {DeVree}, \citenamefont {Zou}, \citenamefont
  {Kruse}, \citenamefont {Chung}, \citenamefont {Kobilka}, \citenamefont
  {Thian}, \citenamefont {Chae}, \citenamefont {Pardon}, \citenamefont
  {Calinski} \emph {et~al.}}]{rasmussen2011crystal}%
  \BibitemOpen
  \bibfield  {author} {\bibinfo {author} {\bibfnamefont {S.~G.}\ \bibnamefont
  {Rasmussen}}, \bibinfo {author} {\bibfnamefont {B.~T.}\ \bibnamefont
  {DeVree}}, \bibinfo {author} {\bibfnamefont {Y.}~\bibnamefont {Zou}},
  \bibinfo {author} {\bibfnamefont {A.~C.}\ \bibnamefont {Kruse}}, \bibinfo
  {author} {\bibfnamefont {K.~Y.}\ \bibnamefont {Chung}}, \bibinfo {author}
  {\bibfnamefont {T.~S.}\ \bibnamefont {Kobilka}}, \bibinfo {author}
  {\bibfnamefont {F.~S.}\ \bibnamefont {Thian}}, \bibinfo {author}
  {\bibfnamefont {P.~S.}\ \bibnamefont {Chae}}, \bibinfo {author}
  {\bibfnamefont {E.}~\bibnamefont {Pardon}}, \bibinfo {author} {\bibfnamefont
  {D.}~\bibnamefont {Calinski}}, \emph {et~al.},\ }\bibfield  {title} {\bibinfo
  {title} {Crystal structure of the $\beta$2 adrenergic receptor--gs protein
  complex},\ }\href@noop {} {\bibfield  {journal} {\bibinfo  {journal}
  {Nature}\ }\textbf {\bibinfo {volume} {477}},\ \bibinfo {pages} {549}
  (\bibinfo {year} {2011})}\BibitemShut {NoStop}%
\bibitem [{\citenamefont {Hekstra}\ \emph {et~al.}(2016)\citenamefont
  {Hekstra}, \citenamefont {White}, \citenamefont {Socolich}, \citenamefont
  {Henning}, \citenamefont {{\v{S}}rajer},\ and\ \citenamefont
  {Ranganathan}}]{hekstra2016electric}%
  \BibitemOpen
  \bibfield  {author} {\bibinfo {author} {\bibfnamefont {D.~R.}\ \bibnamefont
  {Hekstra}}, \bibinfo {author} {\bibfnamefont {K.~I.}\ \bibnamefont {White}},
  \bibinfo {author} {\bibfnamefont {M.~A.}\ \bibnamefont {Socolich}}, \bibinfo
  {author} {\bibfnamefont {R.~W.}\ \bibnamefont {Henning}}, \bibinfo {author}
  {\bibfnamefont {V.}~\bibnamefont {{\v{S}}rajer}},\ and\ \bibinfo {author}
  {\bibfnamefont {R.}~\bibnamefont {Ranganathan}},\ }\bibfield  {title}
  {\bibinfo {title} {Electric-field-stimulated protein mechanics},\ }\href@noop
  {} {\bibfield  {journal} {\bibinfo  {journal} {Nature}\ }\textbf {\bibinfo
  {volume} {540}},\ \bibinfo {pages} {400} (\bibinfo {year}
  {2016})}\BibitemShut {NoStop}%
\bibitem [{\citenamefont {Griffith}\ \emph {et~al.}(2018)\citenamefont
  {Griffith}, \citenamefont {Wiaderek}, \citenamefont {Cibin}, \citenamefont
  {Marbella},\ and\ \citenamefont {Grey}}]{griffith2018niobium}%
  \BibitemOpen
  \bibfield  {author} {\bibinfo {author} {\bibfnamefont {K.~J.}\ \bibnamefont
  {Griffith}}, \bibinfo {author} {\bibfnamefont {K.~M.}\ \bibnamefont
  {Wiaderek}}, \bibinfo {author} {\bibfnamefont {G.}~\bibnamefont {Cibin}},
  \bibinfo {author} {\bibfnamefont {L.~E.}\ \bibnamefont {Marbella}},\ and\
  \bibinfo {author} {\bibfnamefont {C.~P.}\ \bibnamefont {Grey}},\ }\bibfield
  {title} {\bibinfo {title} {Niobium tungsten oxides for high-rate lithium-ion
  energy storage},\ }\href@noop {} {\bibfield  {journal} {\bibinfo  {journal}
  {Nature}\ }\textbf {\bibinfo {volume} {559}},\ \bibinfo {pages} {556}
  (\bibinfo {year} {2018})}\BibitemShut {NoStop}%
\bibitem [{\citenamefont {Cunningham}\ \emph {et~al.}(2019)\citenamefont
  {Cunningham}, \citenamefont {Zhao}, \citenamefont {Parab}, \citenamefont
  {Kantzos}, \citenamefont {Pauza}, \citenamefont {Fezzaa}, \citenamefont
  {Sun},\ and\ \citenamefont {Rollett}}]{cunningham2019keyhole}%
  \BibitemOpen
  \bibfield  {author} {\bibinfo {author} {\bibfnamefont {R.}~\bibnamefont
  {Cunningham}}, \bibinfo {author} {\bibfnamefont {C.}~\bibnamefont {Zhao}},
  \bibinfo {author} {\bibfnamefont {N.}~\bibnamefont {Parab}}, \bibinfo
  {author} {\bibfnamefont {C.}~\bibnamefont {Kantzos}}, \bibinfo {author}
  {\bibfnamefont {J.}~\bibnamefont {Pauza}}, \bibinfo {author} {\bibfnamefont
  {K.}~\bibnamefont {Fezzaa}}, \bibinfo {author} {\bibfnamefont
  {T.}~\bibnamefont {Sun}},\ and\ \bibinfo {author} {\bibfnamefont {A.~D.}\
  \bibnamefont {Rollett}},\ }\bibfield  {title} {\bibinfo {title} {Keyhole
  threshold and morphology in laser melting revealed by ultrahigh-speed x-ray
  imaging},\ }\href@noop {} {\bibfield  {journal} {\bibinfo  {journal}
  {Science}\ }\textbf {\bibinfo {volume} {363}},\ \bibinfo {pages} {849}
  (\bibinfo {year} {2019})}\BibitemShut {NoStop}%
\bibitem [{\citenamefont {Sun}\ \emph {et~al.}(2012)\citenamefont {Sun},
  \citenamefont {Jiang}, \citenamefont {Strzalka}, \citenamefont {Ocola},\ and\
  \citenamefont {Wang}}]{sun2012three}%
  \BibitemOpen
  \bibfield  {author} {\bibinfo {author} {\bibfnamefont {T.}~\bibnamefont
  {Sun}}, \bibinfo {author} {\bibfnamefont {Z.}~\bibnamefont {Jiang}}, \bibinfo
  {author} {\bibfnamefont {J.}~\bibnamefont {Strzalka}}, \bibinfo {author}
  {\bibfnamefont {L.}~\bibnamefont {Ocola}},\ and\ \bibinfo {author}
  {\bibfnamefont {J.}~\bibnamefont {Wang}},\ }\bibfield  {title} {\bibinfo
  {title} {Three-dimensional coherent x-ray surface scattering imaging near
  total external reflection},\ }\href@noop {} {\bibfield  {journal} {\bibinfo
  {journal} {Nature Photonics}\ }\textbf {\bibinfo {volume} {6}},\ \bibinfo
  {pages} {586} (\bibinfo {year} {2012})}\BibitemShut {NoStop}%
\bibitem [{\citenamefont {MacPhee}\ \emph {et~al.}(2002)\citenamefont
  {MacPhee}, \citenamefont {Tate}, \citenamefont {Powell}, \citenamefont {Yue},
  \citenamefont {Renzi}, \citenamefont {Ercan}, \citenamefont {Narayanan},
  \citenamefont {Fontes}, \citenamefont {Walther}, \citenamefont {Schaller}
  \emph {et~al.}}]{macphee2002x}%
  \BibitemOpen
  \bibfield  {author} {\bibinfo {author} {\bibfnamefont {A.~G.}\ \bibnamefont
  {MacPhee}}, \bibinfo {author} {\bibfnamefont {M.~W.}\ \bibnamefont {Tate}},
  \bibinfo {author} {\bibfnamefont {C.~F.}\ \bibnamefont {Powell}}, \bibinfo
  {author} {\bibfnamefont {Y.}~\bibnamefont {Yue}}, \bibinfo {author}
  {\bibfnamefont {M.~J.}\ \bibnamefont {Renzi}}, \bibinfo {author}
  {\bibfnamefont {A.}~\bibnamefont {Ercan}}, \bibinfo {author} {\bibfnamefont
  {S.}~\bibnamefont {Narayanan}}, \bibinfo {author} {\bibfnamefont
  {E.}~\bibnamefont {Fontes}}, \bibinfo {author} {\bibfnamefont
  {J.}~\bibnamefont {Walther}}, \bibinfo {author} {\bibfnamefont
  {J.}~\bibnamefont {Schaller}}, \emph {et~al.},\ }\bibfield  {title} {\bibinfo
  {title} {X-ray imaging of shock waves generated by high-pressure fuel
  sprays},\ }\href@noop {} {\bibfield  {journal} {\bibinfo  {journal}
  {Science}\ }\textbf {\bibinfo {volume} {295}},\ \bibinfo {pages} {1261}
  (\bibinfo {year} {2002})}\BibitemShut {NoStop}%
\bibitem [{\citenamefont {Gr{\'e}aux}\ \emph {et~al.}(2019)\citenamefont
  {Gr{\'e}aux}, \citenamefont {Irifune}, \citenamefont {Higo}, \citenamefont
  {Tange}, \citenamefont {Arimoto}, \citenamefont {Liu},\ and\ \citenamefont
  {Yamada}}]{greaux2019sound}%
  \BibitemOpen
  \bibfield  {author} {\bibinfo {author} {\bibfnamefont {S.}~\bibnamefont
  {Gr{\'e}aux}}, \bibinfo {author} {\bibfnamefont {T.}~\bibnamefont {Irifune}},
  \bibinfo {author} {\bibfnamefont {Y.}~\bibnamefont {Higo}}, \bibinfo {author}
  {\bibfnamefont {Y.}~\bibnamefont {Tange}}, \bibinfo {author} {\bibfnamefont
  {T.}~\bibnamefont {Arimoto}}, \bibinfo {author} {\bibfnamefont
  {Z.}~\bibnamefont {Liu}},\ and\ \bibinfo {author} {\bibfnamefont
  {A.}~\bibnamefont {Yamada}},\ }\bibfield  {title} {\bibinfo {title} {Sound
  velocity of casio3 perovskite suggests the presence of basaltic crust in the
  earth’s lower mantle},\ }\href@noop {} {\bibfield  {journal} {\bibinfo
  {journal} {Nature}\ }\textbf {\bibinfo {volume} {565}},\ \bibinfo {pages}
  {218} (\bibinfo {year} {2019})}\BibitemShut {NoStop}%
\bibitem [{\citenamefont {Chanyshev}\ \emph {et~al.}(2022)\citenamefont
  {Chanyshev}, \citenamefont {Ishii}, \citenamefont {Bondar}, \citenamefont
  {Bhat}, \citenamefont {Kim}, \citenamefont {Farla}, \citenamefont {Nishida},
  \citenamefont {Liu}, \citenamefont {Wang}, \citenamefont {Nakajima} \emph
  {et~al.}}]{chanyshev2022depressed}%
  \BibitemOpen
  \bibfield  {author} {\bibinfo {author} {\bibfnamefont {A.}~\bibnamefont
  {Chanyshev}}, \bibinfo {author} {\bibfnamefont {T.}~\bibnamefont {Ishii}},
  \bibinfo {author} {\bibfnamefont {D.}~\bibnamefont {Bondar}}, \bibinfo
  {author} {\bibfnamefont {S.}~\bibnamefont {Bhat}}, \bibinfo {author}
  {\bibfnamefont {E.~J.}\ \bibnamefont {Kim}}, \bibinfo {author} {\bibfnamefont
  {R.}~\bibnamefont {Farla}}, \bibinfo {author} {\bibfnamefont
  {K.}~\bibnamefont {Nishida}}, \bibinfo {author} {\bibfnamefont
  {Z.}~\bibnamefont {Liu}}, \bibinfo {author} {\bibfnamefont {L.}~\bibnamefont
  {Wang}}, \bibinfo {author} {\bibfnamefont {A.}~\bibnamefont {Nakajima}},
  \emph {et~al.},\ }\bibfield  {title} {\bibinfo {title} {Depressed 660-km
  discontinuity caused by akimotoite--bridgmanite transition},\ }\href@noop {}
  {\bibfield  {journal} {\bibinfo  {journal} {Nature}\ }\textbf {\bibinfo
  {volume} {601}},\ \bibinfo {pages} {69} (\bibinfo {year} {2022})}\BibitemShut
  {NoStop}%
\bibitem [{\citenamefont {Willmott}(2019)}]{willmott2019introduction}%
  \BibitemOpen
  \bibfield  {author} {\bibinfo {author} {\bibfnamefont {P.}~\bibnamefont
  {Willmott}},\ }\href@noop {} {\emph {\bibinfo {title} {An introduction to
  synchrotron radiation: techniques and applications}}}\ (\bibinfo  {publisher}
  {John Wiley \& Sons},\ \bibinfo {year} {2019})\BibitemShut {NoStop}%
\bibitem [{\citenamefont {Deng}\ \emph {et~al.}(2021)\citenamefont {Deng},
  \citenamefont {Chao}, \citenamefont {Feikes}, \citenamefont {Hoehl},
  \citenamefont {Huang}, \citenamefont {Klein}, \citenamefont {Kruschinski},
  \citenamefont {Li}, \citenamefont {Matveenko}, \citenamefont {Petenev} \emph
  {et~al.}}]{deng2021experimental}%
  \BibitemOpen
  \bibfield  {author} {\bibinfo {author} {\bibfnamefont {X.}~\bibnamefont
  {Deng}}, \bibinfo {author} {\bibfnamefont {A.}~\bibnamefont {Chao}}, \bibinfo
  {author} {\bibfnamefont {J.}~\bibnamefont {Feikes}}, \bibinfo {author}
  {\bibfnamefont {A.}~\bibnamefont {Hoehl}}, \bibinfo {author} {\bibfnamefont
  {W.}~\bibnamefont {Huang}}, \bibinfo {author} {\bibfnamefont
  {R.}~\bibnamefont {Klein}}, \bibinfo {author} {\bibfnamefont
  {A.}~\bibnamefont {Kruschinski}}, \bibinfo {author} {\bibfnamefont
  {J.}~\bibnamefont {Li}}, \bibinfo {author} {\bibfnamefont {A.}~\bibnamefont
  {Matveenko}}, \bibinfo {author} {\bibfnamefont {Y.}~\bibnamefont {Petenev}},
  \emph {et~al.},\ }\bibfield  {title} {\bibinfo {title} {Experimental
  demonstration of the mechanism of steady-state microbunching},\ }\href@noop
  {} {\bibfield  {journal} {\bibinfo  {journal} {Nature}\ }\textbf {\bibinfo
  {volume} {590}},\ \bibinfo {pages} {576} (\bibinfo {year}
  {2021})}\BibitemShut {NoStop}%
\bibitem [{\citenamefont {Schoenlein}\ \emph {et~al.}(2000)\citenamefont
  {Schoenlein}, \citenamefont {Chattopadhyay}, \citenamefont {Chong},
  \citenamefont {Glover}, \citenamefont {Heimann}, \citenamefont {Shank},
  \citenamefont {Zholents},\ and\ \citenamefont
  {Zolotorev}}]{schoenlein2000generation}%
  \BibitemOpen
  \bibfield  {author} {\bibinfo {author} {\bibfnamefont {R.}~\bibnamefont
  {Schoenlein}}, \bibinfo {author} {\bibfnamefont {S.}~\bibnamefont
  {Chattopadhyay}}, \bibinfo {author} {\bibfnamefont {H.}~\bibnamefont
  {Chong}}, \bibinfo {author} {\bibfnamefont {T.}~\bibnamefont {Glover}},
  \bibinfo {author} {\bibfnamefont {P.}~\bibnamefont {Heimann}}, \bibinfo
  {author} {\bibfnamefont {C.}~\bibnamefont {Shank}}, \bibinfo {author}
  {\bibfnamefont {A.}~\bibnamefont {Zholents}},\ and\ \bibinfo {author}
  {\bibfnamefont {M.}~\bibnamefont {Zolotorev}},\ }\bibfield  {title} {\bibinfo
  {title} {Generation of femtosecond pulses of synchrotron radiation},\
  }\href@noop {} {\bibfield  {journal} {\bibinfo  {journal} {Science}\ }\textbf
  {\bibinfo {volume} {287}},\ \bibinfo {pages} {2237} (\bibinfo {year}
  {2000})}\BibitemShut {NoStop}%
\bibitem [{\citenamefont {Beaud}\ \emph {et~al.}(2007)\citenamefont {Beaud},
  \citenamefont {Johnson}, \citenamefont {Streun}, \citenamefont {Abela},
  \citenamefont {Abramsohn}, \citenamefont {Grolimund}, \citenamefont
  {Krasniqi}, \citenamefont {Schmidt}, \citenamefont {Schlott},\ and\
  \citenamefont {Ingold}}]{beaud2007spatiotemporal}%
  \BibitemOpen
  \bibfield  {author} {\bibinfo {author} {\bibfnamefont {P.}~\bibnamefont
  {Beaud}}, \bibinfo {author} {\bibfnamefont {S.}~\bibnamefont {Johnson}},
  \bibinfo {author} {\bibfnamefont {A.}~\bibnamefont {Streun}}, \bibinfo
  {author} {\bibfnamefont {R.}~\bibnamefont {Abela}}, \bibinfo {author}
  {\bibfnamefont {D.}~\bibnamefont {Abramsohn}}, \bibinfo {author}
  {\bibfnamefont {D.}~\bibnamefont {Grolimund}}, \bibinfo {author}
  {\bibfnamefont {F.}~\bibnamefont {Krasniqi}}, \bibinfo {author}
  {\bibfnamefont {T.}~\bibnamefont {Schmidt}}, \bibinfo {author} {\bibfnamefont
  {V.}~\bibnamefont {Schlott}},\ and\ \bibinfo {author} {\bibfnamefont
  {G.}~\bibnamefont {Ingold}},\ }\bibfield  {title} {\bibinfo {title}
  {Spatiotemporal stability of a femtosecond hard--x-ray undulator source
  studied by control of coherent optical phonons},\ }\href@noop {} {\bibfield
  {journal} {\bibinfo  {journal} {Physical review letters}\ }\textbf {\bibinfo
  {volume} {99}},\ \bibinfo {pages} {174801} (\bibinfo {year}
  {2007})}\BibitemShut {NoStop}%
\bibitem [{\citenamefont {Zholents}\ and\ \citenamefont
  {Zolotorev}(1996)}]{zholents1996femtosecond}%
  \BibitemOpen
  \bibfield  {author} {\bibinfo {author} {\bibfnamefont {A.}~\bibnamefont
  {Zholents}}\ and\ \bibinfo {author} {\bibfnamefont {M.}~\bibnamefont
  {Zolotorev}},\ }\bibfield  {title} {\bibinfo {title} {Femtosecond x-ray
  pulses of synchrotron radiation},\ }\href@noop {} {\bibfield  {journal}
  {\bibinfo  {journal} {Physical Review Letters}\ }\textbf {\bibinfo {volume}
  {76}},\ \bibinfo {pages} {912} (\bibinfo {year} {1996})}\BibitemShut
  {NoStop}%
\bibitem [{\citenamefont {Holldack}\ \emph {et~al.}(2014)\citenamefont
  {Holldack}, \citenamefont {Ovsyannikov}, \citenamefont {Kuske}, \citenamefont
  {M{\"u}ller}, \citenamefont {Sch{\"a}licke}, \citenamefont {Scheer},
  \citenamefont {Gorgoi}, \citenamefont {K{\"u}hn}, \citenamefont {Leitner},
  \citenamefont {Svensson} \emph {et~al.}}]{holldack2014single}%
  \BibitemOpen
  \bibfield  {author} {\bibinfo {author} {\bibfnamefont {K.}~\bibnamefont
  {Holldack}}, \bibinfo {author} {\bibfnamefont {R.}~\bibnamefont
  {Ovsyannikov}}, \bibinfo {author} {\bibfnamefont {P.}~\bibnamefont {Kuske}},
  \bibinfo {author} {\bibfnamefont {R.}~\bibnamefont {M{\"u}ller}}, \bibinfo
  {author} {\bibfnamefont {A.}~\bibnamefont {Sch{\"a}licke}}, \bibinfo {author}
  {\bibfnamefont {M.}~\bibnamefont {Scheer}}, \bibinfo {author} {\bibfnamefont
  {M.}~\bibnamefont {Gorgoi}}, \bibinfo {author} {\bibfnamefont
  {D.}~\bibnamefont {K{\"u}hn}}, \bibinfo {author} {\bibfnamefont
  {T.}~\bibnamefont {Leitner}}, \bibinfo {author} {\bibfnamefont
  {S.}~\bibnamefont {Svensson}}, \emph {et~al.},\ }\bibfield  {title} {\bibinfo
  {title} {Single bunch x-ray pulses on demand from a multi-bunch synchrotron
  radiation source},\ }\href@noop {} {\bibfield  {journal} {\bibinfo  {journal}
  {Nature Communications}\ }\textbf {\bibinfo {volume} {5}},\ \bibinfo {pages}
  {1} (\bibinfo {year} {2014})}\BibitemShut {NoStop}%
\bibitem [{\citenamefont {Martin}\ \emph {et~al.}(2011)\citenamefont {Martin},
  \citenamefont {Rehm}, \citenamefont {Thomas},\ and\ \citenamefont
  {Bartolini}}]{martin2011experience}%
  \BibitemOpen
  \bibfield  {author} {\bibinfo {author} {\bibfnamefont {I.}~\bibnamefont
  {Martin}}, \bibinfo {author} {\bibfnamefont {G.}~\bibnamefont {Rehm}},
  \bibinfo {author} {\bibfnamefont {C.}~\bibnamefont {Thomas}},\ and\ \bibinfo
  {author} {\bibfnamefont {R.}~\bibnamefont {Bartolini}},\ }\bibfield  {title}
  {\bibinfo {title} {Experience with low-alpha lattices at the diamond light
  source},\ }\href@noop {} {\bibfield  {journal} {\bibinfo  {journal} {Physical
  Review Special Topics-Accelerators and Beams}\ }\textbf {\bibinfo {volume}
  {14}},\ \bibinfo {pages} {040705} (\bibinfo {year} {2011})}\BibitemShut
  {NoStop}%
\bibitem [{\citenamefont {F{\"o}rster}\ \emph {et~al.}(2015)\citenamefont
  {F{\"o}rster}, \citenamefont {Lindenau}, \citenamefont {Leyendecker},
  \citenamefont {Janssen}, \citenamefont {Winkler}, \citenamefont {Schumann},
  \citenamefont {Kirschner}, \citenamefont {Holldack},\ and\ \citenamefont
  {F{\"o}hlisch}}]{forster2015phase}%
  \BibitemOpen
  \bibfield  {author} {\bibinfo {author} {\bibfnamefont {D.~F.}\ \bibnamefont
  {F{\"o}rster}}, \bibinfo {author} {\bibfnamefont {B.}~\bibnamefont
  {Lindenau}}, \bibinfo {author} {\bibfnamefont {M.}~\bibnamefont
  {Leyendecker}}, \bibinfo {author} {\bibfnamefont {F.}~\bibnamefont
  {Janssen}}, \bibinfo {author} {\bibfnamefont {C.}~\bibnamefont {Winkler}},
  \bibinfo {author} {\bibfnamefont {F.~O.}\ \bibnamefont {Schumann}}, \bibinfo
  {author} {\bibfnamefont {J.}~\bibnamefont {Kirschner}}, \bibinfo {author}
  {\bibfnamefont {K.}~\bibnamefont {Holldack}},\ and\ \bibinfo {author}
  {\bibfnamefont {A.}~\bibnamefont {F{\"o}hlisch}},\ }\bibfield  {title}
  {\bibinfo {title} {Phase-locked mhz pulse selector for x-ray sources},\
  }\href@noop {} {\bibfield  {journal} {\bibinfo  {journal} {Optics letters}\
  }\textbf {\bibinfo {volume} {40}},\ \bibinfo {pages} {2265} (\bibinfo {year}
  {2015})}\BibitemShut {NoStop}%
\bibitem [{\citenamefont {Osawa}\ \emph {et~al.}(2017)\citenamefont {Osawa},
  \citenamefont {Ohkochi}, \citenamefont {Fujisawa}, \citenamefont {Kimura},\
  and\ \citenamefont {Kinoshita}}]{osawa2017development}%
  \BibitemOpen
  \bibfield  {author} {\bibinfo {author} {\bibfnamefont {H.}~\bibnamefont
  {Osawa}}, \bibinfo {author} {\bibfnamefont {T.}~\bibnamefont {Ohkochi}},
  \bibinfo {author} {\bibfnamefont {M.}~\bibnamefont {Fujisawa}}, \bibinfo
  {author} {\bibfnamefont {S.}~\bibnamefont {Kimura}},\ and\ \bibinfo {author}
  {\bibfnamefont {T.}~\bibnamefont {Kinoshita}},\ }\bibfield  {title} {\bibinfo
  {title} {Development of optical choppers for time-resolved measurements at
  soft x-ray synchrotron radiation beamlines},\ }\href@noop {} {\bibfield
  {journal} {\bibinfo  {journal} {Journal of Synchrotron Radiation}\ }\textbf
  {\bibinfo {volume} {24}},\ \bibinfo {pages} {560} (\bibinfo {year}
  {2017})}\BibitemShut {NoStop}%
\bibitem [{\citenamefont {Grigoriev}\ \emph {et~al.}(2006)\citenamefont
  {Grigoriev}, \citenamefont {Do}, \citenamefont {Kim}, \citenamefont {Eom},
  \citenamefont {Evans}, \citenamefont {Adams},\ and\ \citenamefont
  {Dufresne}}]{grigoriev2006subnanosecond}%
  \BibitemOpen
  \bibfield  {author} {\bibinfo {author} {\bibfnamefont {A.}~\bibnamefont
  {Grigoriev}}, \bibinfo {author} {\bibfnamefont {D.-H.}\ \bibnamefont {Do}},
  \bibinfo {author} {\bibfnamefont {D.~M.}\ \bibnamefont {Kim}}, \bibinfo
  {author} {\bibfnamefont {C.-B.}\ \bibnamefont {Eom}}, \bibinfo {author}
  {\bibfnamefont {P.~G.}\ \bibnamefont {Evans}}, \bibinfo {author}
  {\bibfnamefont {B.}~\bibnamefont {Adams}},\ and\ \bibinfo {author}
  {\bibfnamefont {E.~M.}\ \bibnamefont {Dufresne}},\ }\bibfield  {title}
  {\bibinfo {title} {Subnanosecond piezoelectric x-ray switch},\ }\href@noop {}
  {\bibfield  {journal} {\bibinfo  {journal} {Applied physics letters}\
  }\textbf {\bibinfo {volume} {89}},\ \bibinfo {pages} {021109} (\bibinfo
  {year} {2006})}\BibitemShut {NoStop}%
\bibitem [{\citenamefont {Larsson}\ \emph {et~al.}(1998)\citenamefont
  {Larsson}, \citenamefont {Heimann}, \citenamefont {Lindenberg}, \citenamefont
  {Schuck}, \citenamefont {Bucksbaum}, \citenamefont {Lee}, \citenamefont
  {Padmore}, \citenamefont {Wark},\ and\ \citenamefont
  {Falcone}}]{larsson1998ultrafast}%
  \BibitemOpen
  \bibfield  {author} {\bibinfo {author} {\bibfnamefont {J.}~\bibnamefont
  {Larsson}}, \bibinfo {author} {\bibfnamefont {P.~A.}\ \bibnamefont
  {Heimann}}, \bibinfo {author} {\bibfnamefont {A.}~\bibnamefont {Lindenberg}},
  \bibinfo {author} {\bibfnamefont {P.}~\bibnamefont {Schuck}}, \bibinfo
  {author} {\bibfnamefont {P.}~\bibnamefont {Bucksbaum}}, \bibinfo {author}
  {\bibfnamefont {R.}~\bibnamefont {Lee}}, \bibinfo {author} {\bibfnamefont
  {H.}~\bibnamefont {Padmore}}, \bibinfo {author} {\bibfnamefont
  {J.}~\bibnamefont {Wark}},\ and\ \bibinfo {author} {\bibfnamefont
  {R.}~\bibnamefont {Falcone}},\ }\bibfield  {title} {\bibinfo {title}
  {Ultrafast structural changes measured by time-resolved x-ray diffraction.},\
  }\href@noop {} {\bibfield  {journal} {\bibinfo  {journal} {Applied Physics A:
  Materials Science \& Processing}\ }\textbf {\bibinfo {volume} {66}} (\bibinfo
  {year} {1998})}\BibitemShut {NoStop}%
\bibitem [{\citenamefont {Gaal}\ \emph {et~al.}(2014)\citenamefont {Gaal},
  \citenamefont {Schick}, \citenamefont {Herzog}, \citenamefont {Bojahr},
  \citenamefont {Shayduk}, \citenamefont {Goldshteyn}, \citenamefont
  {Leitenberger}, \citenamefont {Vrejoiu}, \citenamefont {Khakhulin},
  \citenamefont {Wulff} \emph {et~al.}}]{gaal2014ultrafast}%
  \BibitemOpen
  \bibfield  {author} {\bibinfo {author} {\bibfnamefont {P.}~\bibnamefont
  {Gaal}}, \bibinfo {author} {\bibfnamefont {D.}~\bibnamefont {Schick}},
  \bibinfo {author} {\bibfnamefont {M.}~\bibnamefont {Herzog}}, \bibinfo
  {author} {\bibfnamefont {A.}~\bibnamefont {Bojahr}}, \bibinfo {author}
  {\bibfnamefont {R.}~\bibnamefont {Shayduk}}, \bibinfo {author} {\bibfnamefont
  {J.}~\bibnamefont {Goldshteyn}}, \bibinfo {author} {\bibfnamefont
  {W.}~\bibnamefont {Leitenberger}}, \bibinfo {author} {\bibfnamefont
  {I.}~\bibnamefont {Vrejoiu}}, \bibinfo {author} {\bibfnamefont
  {D.}~\bibnamefont {Khakhulin}}, \bibinfo {author} {\bibfnamefont
  {M.}~\bibnamefont {Wulff}}, \emph {et~al.},\ }\bibfield  {title} {\bibinfo
  {title} {Ultrafast switching of hard x-rays},\ }\href@noop {} {\bibfield
  {journal} {\bibinfo  {journal} {Journal of synchrotron radiation}\ }\textbf
  {\bibinfo {volume} {21}},\ \bibinfo {pages} {380} (\bibinfo {year}
  {2014})}\BibitemShut {NoStop}%
\bibitem [{\citenamefont {Chen}\ \emph {et~al.}(2019)\citenamefont {Chen},
  \citenamefont {Jung}, \citenamefont {Walko}, \citenamefont {Li},
  \citenamefont {Gao}, \citenamefont {Shenoy}, \citenamefont {L{\'o}pez},\ and\
  \citenamefont {Wang}}]{chen2019ultrafast}%
  \BibitemOpen
  \bibfield  {author} {\bibinfo {author} {\bibfnamefont {P.}~\bibnamefont
  {Chen}}, \bibinfo {author} {\bibfnamefont {I.~W.}\ \bibnamefont {Jung}},
  \bibinfo {author} {\bibfnamefont {D.~A.}\ \bibnamefont {Walko}}, \bibinfo
  {author} {\bibfnamefont {Z.}~\bibnamefont {Li}}, \bibinfo {author}
  {\bibfnamefont {Y.}~\bibnamefont {Gao}}, \bibinfo {author} {\bibfnamefont
  {G.~K.}\ \bibnamefont {Shenoy}}, \bibinfo {author} {\bibfnamefont
  {D.}~\bibnamefont {L{\'o}pez}},\ and\ \bibinfo {author} {\bibfnamefont
  {J.}~\bibnamefont {Wang}},\ }\bibfield  {title} {\bibinfo {title} {Ultrafast
  photonic micro-systems to manipulate hard x-rays at 300 picoseconds},\
  }\href@noop {} {\bibfield  {journal} {\bibinfo  {journal} {Nature
  communications}\ }\textbf {\bibinfo {volume} {10}},\ \bibinfo {pages} {1}
  (\bibinfo {year} {2019})}\BibitemShut {NoStop}%
\bibitem [{\citenamefont {Antonio}\ \emph {et~al.}(2012)\citenamefont
  {Antonio}, \citenamefont {Zanette},\ and\ \citenamefont
  {L{\'o}pez}}]{antonio2012frequency}%
  \BibitemOpen
  \bibfield  {author} {\bibinfo {author} {\bibfnamefont {D.}~\bibnamefont
  {Antonio}}, \bibinfo {author} {\bibfnamefont {D.~H.}\ \bibnamefont
  {Zanette}},\ and\ \bibinfo {author} {\bibfnamefont {D.}~\bibnamefont
  {L{\'o}pez}},\ }\bibfield  {title} {\bibinfo {title} {Frequency stabilization
  in nonlinear micromechanical oscillators},\ }\href@noop {} {\bibfield
  {journal} {\bibinfo  {journal} {Nature communications}\ }\textbf {\bibinfo
  {volume} {3}},\ \bibinfo {pages} {1} (\bibinfo {year} {2012})}\BibitemShut
  {NoStop}%
\bibitem [{\citenamefont {Zhou}\ \emph {et~al.}(2020)\citenamefont {Zhou},
  \citenamefont {Moldovan}, \citenamefont {Stan}, \citenamefont {Cai},
  \citenamefont {Czaplewski},\ and\ \citenamefont
  {L{\'o}pez}}]{zhou2020approaching}%
  \BibitemOpen
  \bibfield  {author} {\bibinfo {author} {\bibfnamefont {J.}~\bibnamefont
  {Zhou}}, \bibinfo {author} {\bibfnamefont {N.}~\bibnamefont {Moldovan}},
  \bibinfo {author} {\bibfnamefont {L.}~\bibnamefont {Stan}}, \bibinfo {author}
  {\bibfnamefont {H.}~\bibnamefont {Cai}}, \bibinfo {author} {\bibfnamefont
  {D.~A.}\ \bibnamefont {Czaplewski}},\ and\ \bibinfo {author} {\bibfnamefont
  {D.}~\bibnamefont {L{\'o}pez}},\ }\bibfield  {title} {\bibinfo {title}
  {Approaching the strain-free limit in ultrathin nanomechanical resonators},\
  }\href@noop {} {\bibfield  {journal} {\bibinfo  {journal} {Nano letters}\
  }\textbf {\bibinfo {volume} {20}},\ \bibinfo {pages} {5693} (\bibinfo {year}
  {2020})}\BibitemShut {NoStop}%
\bibitem [{\citenamefont {Schmid}\ \emph {et~al.}(2016)\citenamefont {Schmid},
  \citenamefont {Villanueva},\ and\ \citenamefont
  {Roukes}}]{schmid2016fundamentals}%
  \BibitemOpen
  \bibfield  {author} {\bibinfo {author} {\bibfnamefont {S.}~\bibnamefont
  {Schmid}}, \bibinfo {author} {\bibfnamefont {L.~G.}\ \bibnamefont
  {Villanueva}},\ and\ \bibinfo {author} {\bibfnamefont {M.~L.}\ \bibnamefont
  {Roukes}},\ }\href@noop {} {\emph {\bibinfo {title} {Fundamentals of
  nanomechanical resonators}}},\ Vol.~\bibinfo {volume} {49}\ (\bibinfo
  {publisher} {Springer},\ \bibinfo {year} {2016})\BibitemShut {NoStop}%
\bibitem [{\citenamefont {Ugural}\ and\ \citenamefont
  {Fenster}(2003)}]{ugural2003advanced}%
  \BibitemOpen
  \bibfield  {author} {\bibinfo {author} {\bibfnamefont {A.~C.}\ \bibnamefont
  {Ugural}}\ and\ \bibinfo {author} {\bibfnamefont {S.~K.}\ \bibnamefont
  {Fenster}},\ }\href@noop {} {\emph {\bibinfo {title} {Advanced strength and
  applied elasticity}}}\ (\bibinfo  {publisher} {Pearson education},\ \bibinfo
  {year} {2003})\BibitemShut {NoStop}%
\end{thebibliography}%

\setcounter{figure}{0} \renewcommand{\thefigure}{S\arabic{figure}}

\begin{figure*}[htp]
	\centering 
	\includegraphics[page=1, trim = 0.7in 2.5in 5.3in 1.7in,  clip, width=0.94\textwidth]{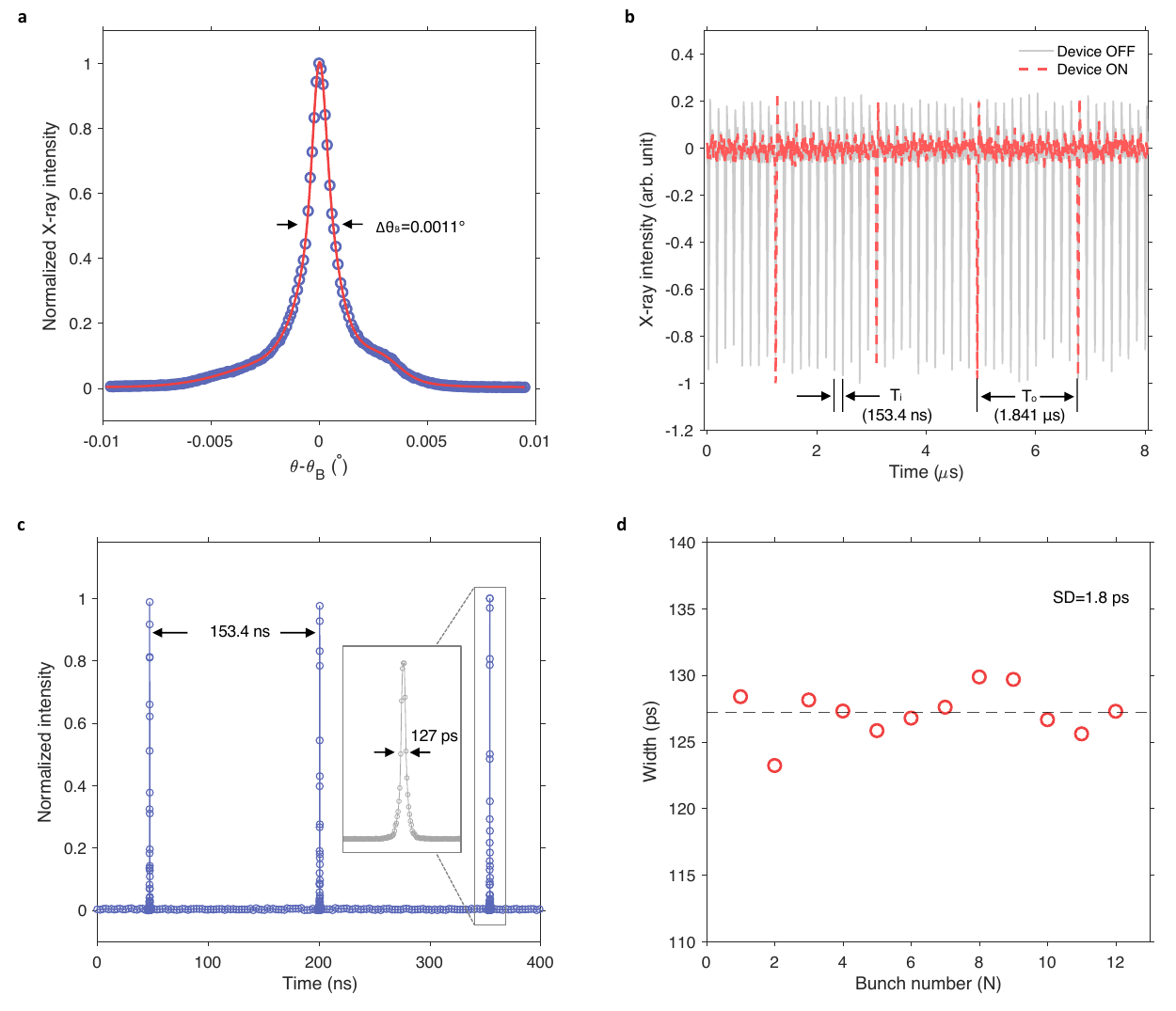}
	\caption{\textbf{$\vert$ Ultrafast X-ray manipulation at APS 24-bunch mode (8 $\textbf{keV}$ energy, 78.7 $\textbf{ps}$ bunch length).} \textbf{a}, Static rocking curve. When the device is static, the diffractive beam intensity was measured by varying the X-ray incident angle in relation to the torsional mirror surface with a precise diffractometer. A prominent Si (400) diffraction peak shows up when the Bragg condition is fulfilled ($\theta_B=34.803^{\circ}$). \textbf{b}, Oscilloscope time traces show the dynamic X-ray bunch modulation ($T_o= 12\;T_i$). The equilibrium position of the shaper surface was aligned at Bragg angle $\theta_B$. \textbf{d} and \textbf{d}, Dynamic rocking curve (c) and bunch width deviations (d) measured with delay time scan at 8.8° torsional amplitude.}
	\label{fig.S01}
\end{figure*}

\begin{figure*}[htp]
	\centering 
	\includegraphics[page=1, trim = 0.7in 4.7in 5.3in 1.7in,  clip, width=0.94\textwidth]{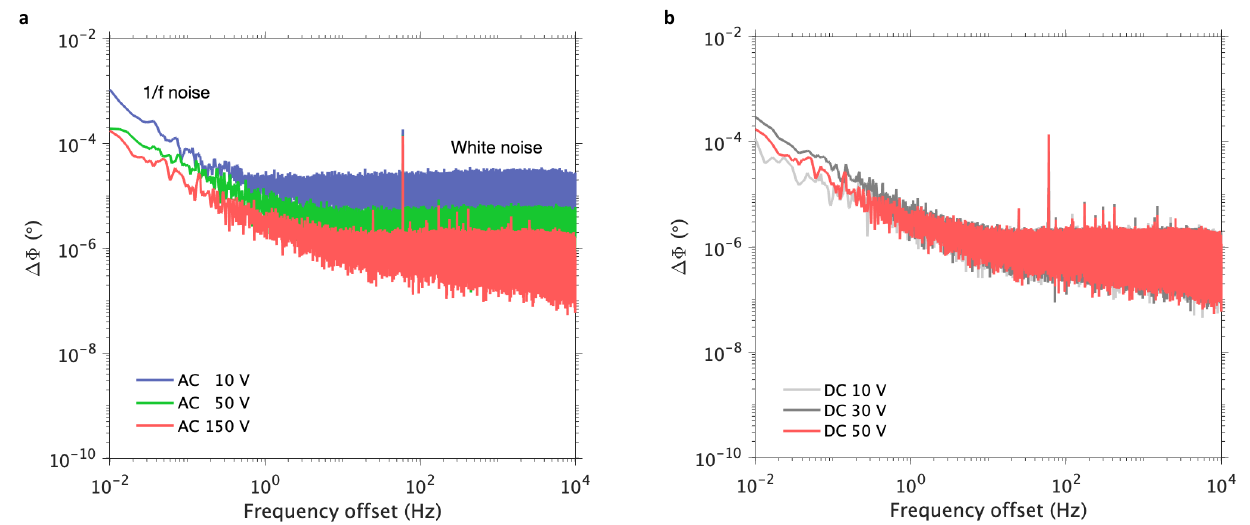}
	\caption{\textbf{$\vert$ System electronic noise}. \textbf{a}, Phase noise measured at 50 $V$ DC bias voltage. \textbf{b}, Phase noise measured at 150 $V$ AC excitation voltage. The system noise is inversely proportional to the AC driven voltage and approximately independent of DC bias voltage.}
	\label{fig.S02}
\end{figure*}

\begin{figure*}[htp]
	\centering 
	\includegraphics[page=1, trim = 0.7in 2.65in 5.2in 1.7in,  clip, width=0.94\textwidth]{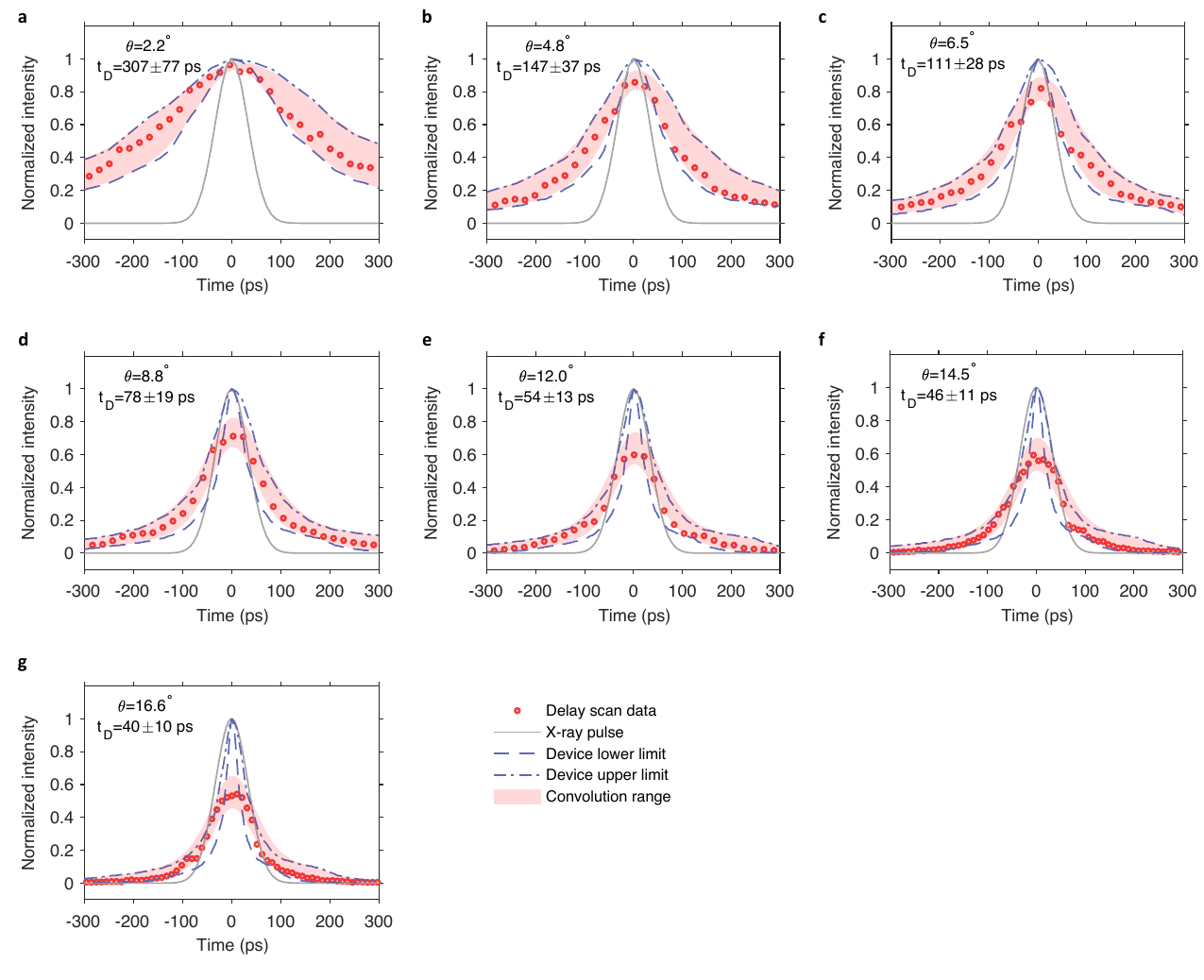}
	\caption{\textbf{$\vert$ Characterization of the shaper diffractive time window from delay time scan}. The dynamic rocking curve measured with delay time scan is the convolution of the X-ray pulse (width 78.7 $ps$) and the shaper diffractive time window (width $t_D$). }
	\label{fig.S03}
\end{figure*}

\begin{figure*}[htp]
	\centering 
	\includegraphics[page=1, trim = 0.7in 3.85in 5.1in 1.7in,  clip, width=0.94\textwidth]{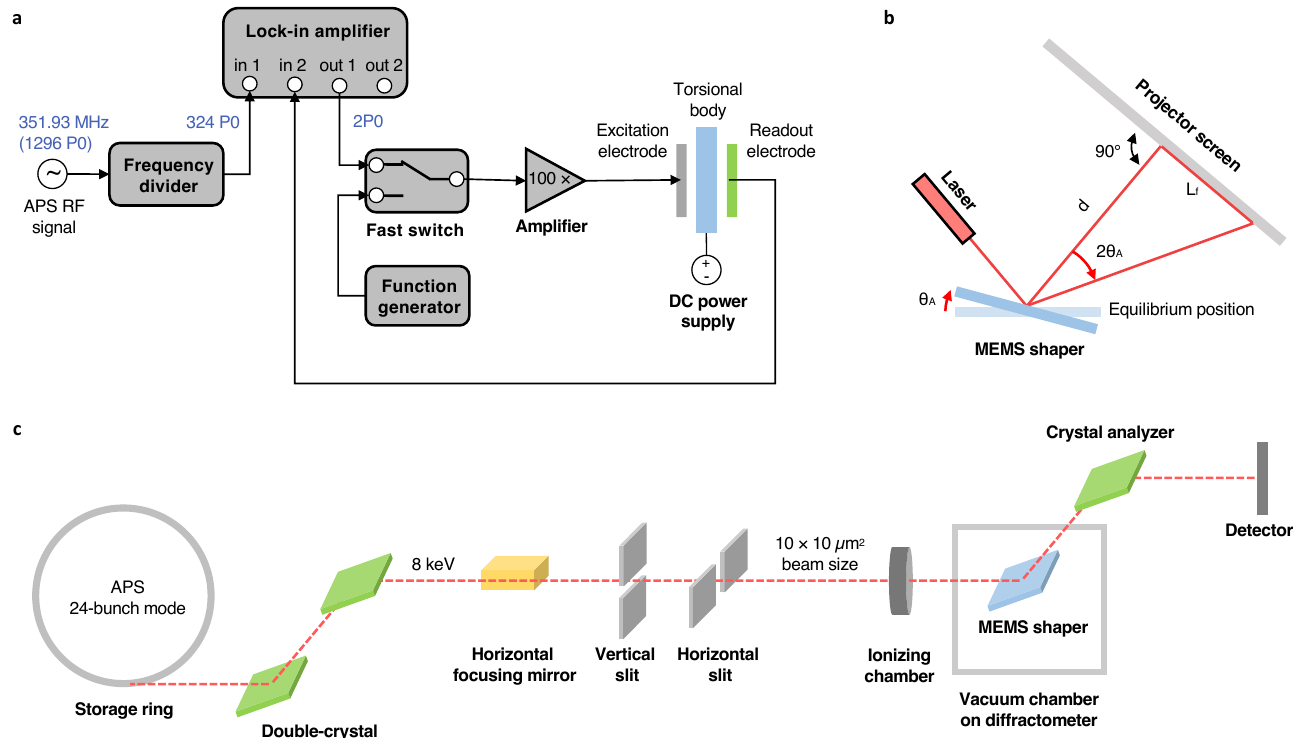}
	\caption{\textbf{$\vert$ Schematic of experimental setup}. \textbf{a}, Schematic of electronic setup. \textbf{b}, Schematic of the setup for the torsional amplitude measurement. \textbf{c}, Schematic of X-ray manipulation setup.}
	\label{fig.S04}
\end{figure*}

\end{document}